\documentclass[%
reprint,
amsmath,amssymb,
aps,pre,
]{revtex4-2}
\usepackage{float}
\usepackage{babel} 
\usepackage[utf8]{inputenc}
\usepackage{lineno,hyperref}
\usepackage{amsbsy}
\usepackage{amstext}
\usepackage{amssymb}
\usepackage{graphicx}
\usepackage{esint}
\usepackage{amsmath}
\usepackage[all]{xy}
\usepackage{multirow,array}
\usepackage{xcolor}
\usepackage{units}
\usepackage{multirow}
\usepackage{array}
\usepackage{rotating}
\usepackage{dcolumn}
\usepackage{bm}
\usepackage{fancyhdr} 
\usepackage{braket}
\usepackage{color} 	
\usepackage{lastpage} 
\usepackage{hyperref} 	
\usepackage{url}
\usepackage{bibentry}
\usepackage{multirow}
\usepackage{silence}
\usepackage{appendix}
\usepackage{natbib}
\usepackage{physics}
\usepackage{ulem}
\usepackage{cleveref}

\begin{document}
\title{Spectral truncation of out-of-time-ordered correlators in dissipative system}

\author{Pablo D. Bergamasco}
\affiliation{Departamento de F\'isica, CNEA, Libertador 8250, (C1429BNP) Buenos Aires, Argentina}
\author{Gabriel G. Carlo}
\author{Alejandro M. F. Rivas}
\affiliation{Departamento de F\'isica, CNEA, CONICET, Libertador 8250, (C1429BNP) Buenos Aires, Argentina}
\date{\today}

\begin{abstract}
  Out-of-time-ordered correlators (OTOCs) have emerged as powerful tools for diagnosing quantum chaos and information scrambling. While extensively studied in closed quantum systems, their behavior in dissipative environments remains less understood. In this work, we investigate the spectral decomposition of OTOCs in open quantum systems, using the dissipative modified kicked rotator (DMKR) as a paradigmatic model. By analyzing the eigenvalue spectrum of the quantum Liouvillian, we identify a crucial spectral truncation criterion that enables efficient modeling of OTOC dynamics. Our results reveal two distinct temporal regimes: a long-time decay phase governed by the spectral gap and an intermediate-time regime where a small subset of subdominant eigenvalues plays a crucial role. This spectral truncation criterion allows for efficient modeling of OTOC decay and reveals a direct connection between eigenvalue structure and information scrambling. Our results provide a quantitative framework for understanding OTOCs in dissipative quantum systems and suggest new avenues for experimental exploration in open quantum platforms. 
\end{abstract}

\maketitle

\section{\label{sec:Intro} Introduction}
The out-of-time-ordered correlator (OTOC), initially conceived in superconductivity to characterize quasiclassical chaotic systems \cite{larkin1969quasiclassical}, has recently ascended to a central position in quantum chaos and quantum information theory. Defined mathematically in \cref{eq:otoc}, the OTOC serves as a crucial diagnostic tool for exploring quantum chaos and information scrambling within closed quantum systems. Its ability to bridge the gap between quantum dynamics and classical chaos, notably through its connection to the classical Lyapunov exponent, provides a unique lens into phenomena such as operator growth and complexity.  Fueled by seminal insights from black hole physics \cite{maldacena2016bound}, the OTOC's applicability has rapidly expanded across a diverse range of fields, encompassing many-body physics\cite{shenker2014black, aleiner2016microscopic,huang2017out,borgonovi2018emergence,slagle2017out,chen2017out,garcia20}, high-energy theory \cite{akutagawa2020out}, and the broader domain of quantum chaos \cite{notenson2023classical, lakshminarayan2018out, wang2021quantum,jalabert2018semiclassical}. In closed systems, the OTOC not only effectively quantifies information scrambling \cite{swingle2018unscrambling} and quantum complexity \cite{PhysRevResearch.1.033044, bergamasco2017,Benenti-Carlo-Prosen,PhysRevResearch.2.043178, prakash2019scrambling} but also excels at distinguishing between chaotic and regular dynamics, often linked to entropic measures and quantum complexity itself \cite{hosur2016chaos, fan2017out,bergamasco2020relevant}.

\begin{gather}
  C(t) = \expval{\left[ A(t) B(0)\right] \left[ A(t) B(0)\right]^{\dagger}}
\label{eq:otoc}
\end{gather}

However, the picture becomes significantly more intricate when considering open quantum systems, where the inevitable presence of decoherence fundamentally alters the interplay with scrambling.  While recent investigations suggest that bipartite OTOCs might offer a pathway to differentiate between chaotic and regular regimes even in dissipative many-body systems \cite{zanardi2021information}, their general applicability and robustness, especially in physically realistic environments \cite{halpern2018}, remain open questions.  Challenging the notion that OTOCs are unable to disentangle scrambling from decoherence \cite{touil2020quantum}, our prior work \cite{bergamasco2023} took a different approach. We explored a paradigmatic dissipative system with a classical counterpart—the dissipative modified kicked rotator (DMKR)—to investigate the OTOC's behavior in a more generic setting. We found that, while lacking short-time exponential growth, the OTOC exhibits a long-time decay ($t > 5$) with a rate that strikingly mirrors the classical Lyapunov exponent.  This decay, dictated by the spectral gap of the quantum Liouvillian ($\hat{\mathcal{L}}$), only aligns with classical predictions upon the introduction of $\hbar_{\text{eff}}$-scale noise, thereby reinforcing the quantum-classical correspondence principle previously highlighted in Refs.~\cite{carlo2019,carlo2012quantum}.

In this work, we advance significantly beyond our initial findings, delving into the spectral underpinnings of OTOC dynamics in dissipative quantum systems. By elucidating the explicit relationship between the Liouvillian eigenvalue spectrum and the OTOC's temporal evolution, we uncover a richer understanding of its behavior. Our analysis reveals two distinct temporal regimes: a long-time decay phase, accurately described by the dominant spectral gap ($\lambda_1$), and a more complex intermediate-time regime ($5 < t < 20$).  Remarkably, we demonstrate that this intermediate regime, while seemingly intricate, can be faithfully reconstructed by considering the contributions from a surprisingly small subset of approximately 10 subdominant Liouvillian eigenvalues. This key observation holds true even across mixed chaotic-regular parameter regions, providing a crucial spectral truncation criterion for efficient OTOC modeling, despite the full Liouvillian spectrum has intractable dimension ($N = 1024$). This finding highlights the crucial role of the Liouvillian eigenvalue spectrum in shaping OTOC dynamics, demonstrating that a small subset of subdominant eigenvalues is sufficient to capture key features of information scrambling and distinguish between regular and chaotic regimes in open quantum systems.

This paper is structured as follows: Section~\ref{sec:system} introduces the dissipative modified kicked rotator model and the numerical methods employed. Section~\ref{sec:otoc} details the spectral decomposition of the OTOC. Section~\ref{sec:results} presents and discusses the numerical results obtained from our spectral analysis. Finally, we conclude with a comprehensive discussion of the implications of our findings for the use of OTOCs as robust quantum chaos diagnostics in open quantum systems.

\section{\label{sec:system} System}
In this work, the same system used in \cite{bergamasco2023} will be employed, which consisted in a Dissipative Modified Kicked Rotator subject to a asymmetric periodic potencial given by,
\begin{equation}
    V(q,t) = k \left[\cos{(q)} + \frac{a}{2}\cos{(2q+\phi)}\right]\sum_{m=-\infty}^{\infty}{\delta(t - m\tau)},
    \label{eq:potencial}
\end{equation}
where $k$ denotes the strength of the kick and $\tau$ its period. Incorporating dissipation results in the following map:
\begin{equation}
    \Bar{n} = \gamma n + k[\sin{(q)} + \sin{(2q+\phi)}]\qquad  \Bar{q} = q + \tau \Bar{n},
    \label{eq:krmap}
\end{equation}
where $n$ ($\Bar{n}$) is the momentum variable conjugate to $q$ ($\Bar{q}$) before (after) the kick and $\gamma$ $(0 \leq \gamma \leq 1)$ is the dissipation parameter. When $\gamma = 1$, the conservative system is recovered. Conversely, setting $\gamma = 0$ corresponds to maximum environmental strength. In this type of systems, it is customary to define the scaled momentum $p = \tau n$ and the quantity $K = \tau k$. The parameters $a = 0.5$ and $b = 0.5$ are selected to yield a rich dynamical landscape, which is suitable for the present investigation. 

The quantum counterpart is obtained by following a standard process: $q\to\hat{q}$ and $n\to\hat{n} = -i(d/dq)$ $(\hbar = 1)$. Under the assumption that $[q,p] = i\tau$ (where $\hat{p}=\hat{n}$), the effective Planck constant is defined by identifying $\hbar_{\text{eff}}=\tau$. In the classical limits, $\hbar_{\text{eff}} \rightarrow 0$ and $K=\hbar_{\text{eff}} k$ remains constant. To incorporate dissipation, we will do so in the usual way, which is through Lindblad's master equation \cite{lindblad1976} to describe the evolution of the operators in the Heisenberg representation.

\begin{align}
\Dot{\hat{B}} = i[\hat{H}_{s}, \hat{B}] - \frac{1}{2}\sum_{\nu=1}^{2}\{\hat{M}^{\dagger}_{\nu}\hat{M}_{\nu}, \hat{B}\} +\sum_{\nu=1}^{2} \hat{M}^{\dagger}_{\nu} \hat{B} \hat{M}_{\nu} \equiv \mathcal{L}^{\dagger}(\hat{B}),
\label{eq:lindblad_adj}
\end{align}
where ${H}_{s}=\hat{n}^{2}/2 + V(\hat{q},t)$ is the Hamiltonian of the system, $\{\cdot ,\cdot \}$ the anticommutator, and $\hat{M}_{\nu}$ are the Lindblad operators defined as \cite{carlo2019}:
\begin{equation}
    \begin{gathered}
        \hat{M}_{1} = g \sum_{n} \sqrt{n+1}\ket{n}\bra{n+1}\\
        \hat{M}_{2} = g \sum_{n} \sqrt{n+1}\ket{-n}\bra{-n-1},
    \end{gathered}
\end{equation}
where $ \ket{n} $ are the momentum states with $ n=0, 1, \dots $ and $ g = \sqrt{-\ln\gamma} $ \cite{graham1985, dittrich1989}.

\section{\label{sec:otoc} Spectral descomposition of OTOC}
As previously stated, the operator evolution is governed by \cref{eq:lindblad_adj}. This equation can be written in a vectorized form using the Choi-Jamiolkowski isomorphism \cite{ChoiJamiolkowski}. 
\begin{align}
    \dv{t}\text{vec}(B) = \hat{\mathcal{L}}^{\dagger} \text{vec}(B)
    \label{eq:lindblad_vec_simplificada}
\end{align}
In the context of a Hilbert space of dimension $N$, the operator $\hat{\mathcal{L}}^{\dagger}$ assumes the form of a $N^2 \times N^2$ matrix, while $\text{vec}(B)$ is a vector of $N^2$ components. The operator $\hat{\mathcal{L}}^{\dagger}$ is defined as follows:
\begin{align}
    \hat{\mathcal{L}}^{\dagger} = &i( I \otimes H - H^{T} \otimes I)+ \sum_{\nu}(\hat{M}_{\nu} \otimes \hat{M}_{\nu}^{\dagger})\nonumber \\
    &- \sum_{\nu}(\frac{1}{2} I\otimes \hat{M}_{\nu}^{\dagger} \hat{M}_{\nu} - \frac{1}{2}(\hat{M}_{\nu}^{\dagger}M_{\nu})^{T} \otimes I)
\end{align}

if $\hat{\mathcal{L}}^{\dagger}$ is time-independent, integration of \cref{eq:lindblad_vec_simplificada} will yield $B$ at time $t$ as follows: 
\begin{align}
    \text{vec}(B_t) = \exp(\hat{\mathcal{L}}^{\dagger}t) \text{vec}(B_0) = (\hat{\Lambda}^{\dagger})^{t} \text{vec}(B_0)  
    \label{eq:evolucion_oper}
\end{align}
where $\hat{\Lambda}^{\dagger} = \exp(\hat{\mathcal{L}}^{\dagger})$. In order to simplify the notation, it is now appropriate to refer to $\text{vec}(B)$ as $\ket{B}\rangle$. 

The operator $\hat{\mathcal{L}}^{\dagger}$ is complex and not hermitian, resulting in the existence of eigenvalues and eigenvectors on both the right and the left. 
\begin{align}
    \hat{\mathcal{L}}^{\dagger} \ket{R_i}\rangle &= \sigma_{i} \ket{R_i}\rangle\\
     \langle\bra{L_i}\hat{\mathcal{L}}^{\dagger}  &= \sigma_{i} \langle\bra{L_i}
\end{align}
where $\ket{R_i}\rangle$  and $\ket{L_i}\rangle$ are the left and right eigenvectors, both column vectors and $\sigma_{i}$ are the eigenvalues. $R_i$ and $L_i$ are, therefore, the matrices associated with the eigenvectors. They satisfy the following orthonormality relationships, 
\begin{align}
    \text{ Tr}(L^{\dagger}_{i}R_{j}) = \text{ Tr}(L_{i}R_{j}) = \text{ Tr}(L^{\dagger}_{i}R^{\dagger}_{j}) = \delta_{ij}
\end{align}

Hence, the spectral decomposition of $\hat{\mathcal{L}}^{\dagger}$ and $\hat{\Lambda}^{\dagger}$ can be expressed as follows: 
\begin{gather}
  \hat{\mathcal{L}}^{\dagger} = \sum_{i} \sigma_{i}r_{i}l^{\dagger}_{i} = \sum_{i} \sigma_{i} \ket{R_{i}}\rangle \langle\bra{L_{i}}\\
  (\hat{\Lambda}^{\dagger})^{t} = \sum_{i}{ (e^{\sigma_{i}})^{t} \ket{R_{i}}\rangle \langle\bra{L_{i}}} = \sum_{i}{\lambda_{i}^{t}\ket{R_{i}}\rangle \langle\bra{L_{i}}} 
  \label{eq:descomp_Lambda}
\end{gather}
Finally, using the \cref{eq:evolucion_oper} together with the \cref{eq:descomp_Lambda}, we can write the 
spectral decomposition of the OTOC, $C(t)$, defined in the \cref{eq:otoc} as (see \cref{ap:spectralOTOC}), 
\begin{gather}
  C(t) = \sum_{ij}{(\lambda_{i}\lambda^{*}_{j})^t b_{i}b^{*}_{j} d_{ij}}
    \label{eq:decomp_otoc}
\end{gather}    
where, 
\begin{gather}
  b_{i} = \text{Tr}\left(L^{\dagger}_{i}B(0) \right)\nonumber \\
  d_{ij} = \text{Tr}\left( \left[ \hat{A}, R_{i} \right] \left[ \hat{A}, R_{j} \right]^{\dagger} \rho_{o}\right)\nonumber
\end{gather}

The spectral development of the OTOC facilitates a more detailed comprehension of the eigenvalues of $\hat{\mathcal{L}}^{\dagger}$ that are responsible for the behavior observed at both short and long timescales. 

\section{\label{sec:results} Results}

In the following, we will determine the importance of each term in \cref{eq:decomp_otoc} for different times. This analysis will be repeated for different system parameters, enabling the identification of the involved eigenvalues in the OTOC decay observed in \cite{bergamasco2023}.

The operators employed in this study will be the same as in \cite{bergamasco2023}, namely $\hat{A} = e^{i\hat{Q}}$ and $\hat{B} = \hat{P}$, where $\hat{Q}$ is the position operator and $\hat{P}$ is the momentum operator. We will use a initial coherent state,$\rho_{o}$, centered at $p_o = 0$ and $q_o = \pi$.

The coefficients $\lambda_{i}$, $a_{i}$, and $d_{ij}$ that integrate each term of the sum in \cref{eq:decomp_otoc} are determined at $t = 0$ and will not change when $t$ varies. However, since time appears as an exponent of the product of the eigenvalues, which are less than or equal to $1$, the weight of each term will decrease as time elapses. On the other hand, the relative relevance between terms of the sum is also affected by the magnitude of the 
coefficients $a_{i}$ and $d_{ij}$ in which the projection of the chosen operators on the basis of the eigenstate of $\hat{\mathcal{L}}^{\dagger}$ comes into play. 

We take the absolute value of each term and normalize it by the sum of its modulus to determine the weight of each term, whether complex or real, i.e.,

\begin{gather}
  p_{ij}(t) = \frac{\left|(\lambda_{i}\lambda^{*}_{j})^t b_{i}b^{*}_{j} d_{ij}\right|}{\sum_{ij}{\left|(\lambda_{i}\lambda^{*}_{j})^t b_{i}b^{*}_{j} d_{ij}\right|}
}
\end{gather}
The complex conjugate eigenvalue pairs, by symmetry, possess identical modulus and thus contribute equally to the OTOC's magnitude.

We numerically computed the $100$ eigenvalues with the largest modulus for the system, along with their corresponding right and left eigenvectors. Among them, a single eigenvalue, denoted as $\lambda_0$, has a modulus of $1$ and corresponds to the system's unique attractor. The system parameters were set to a dimension of $N = 1024$, an effective Planck’s constant $h_{\text{eff}} = 0.031$, and a dissipation rate $\gamma = 0.2$, with eigenvalues computed using the Arnoldi method. Given the large dimension of the operator $\hat{\mathcal{L}}$, computing the full spectrum proved computationally prohibitive. 

\begin{figure}[htp]
  \begin{center}
    \includegraphics[width=\columnwidth]{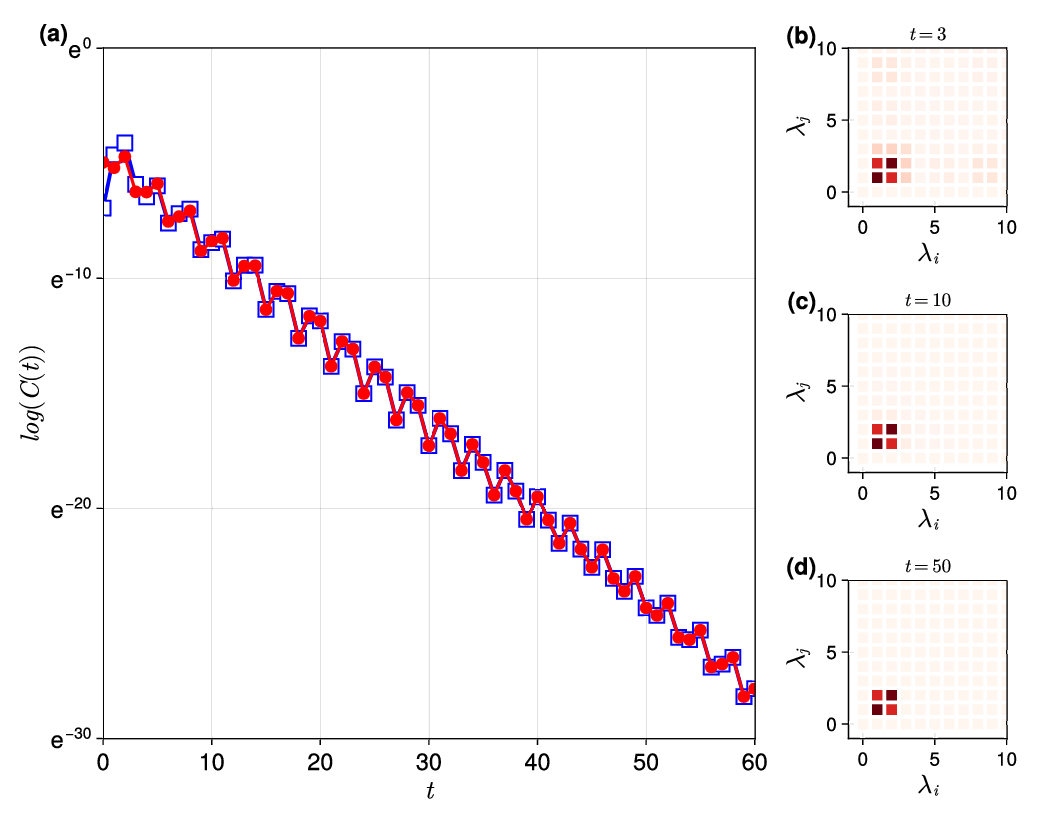}
  \end{center}
  \caption{\textbf{(Color online)} OTOC dynamics and $p_{ij}$ evolution. Panel (A) compares OTOC decay from numerical simulations of \cref{eq:otoc} (blue empty square) with spectral expansion considering only $\lambda_{1}$ and $\lambda_{2}$ (red solid circles) for $K=2.0$. Panels (B)-(D) are heatmaps of $p_{ij}$ magnitude at times $t=3, 10, 50$ respectively, with deeper red/gray indicating larger values. Calculations for Hilbert space dimension $N=1024$, effective Planck constant $h_{eff} = 0.031$ and dissipation rate $\gamma = 0.2$.}
  \label{fig:fig1}
\end{figure}

\begin{figure}[htp]
  \begin{center}
    \includegraphics[width=\columnwidth]{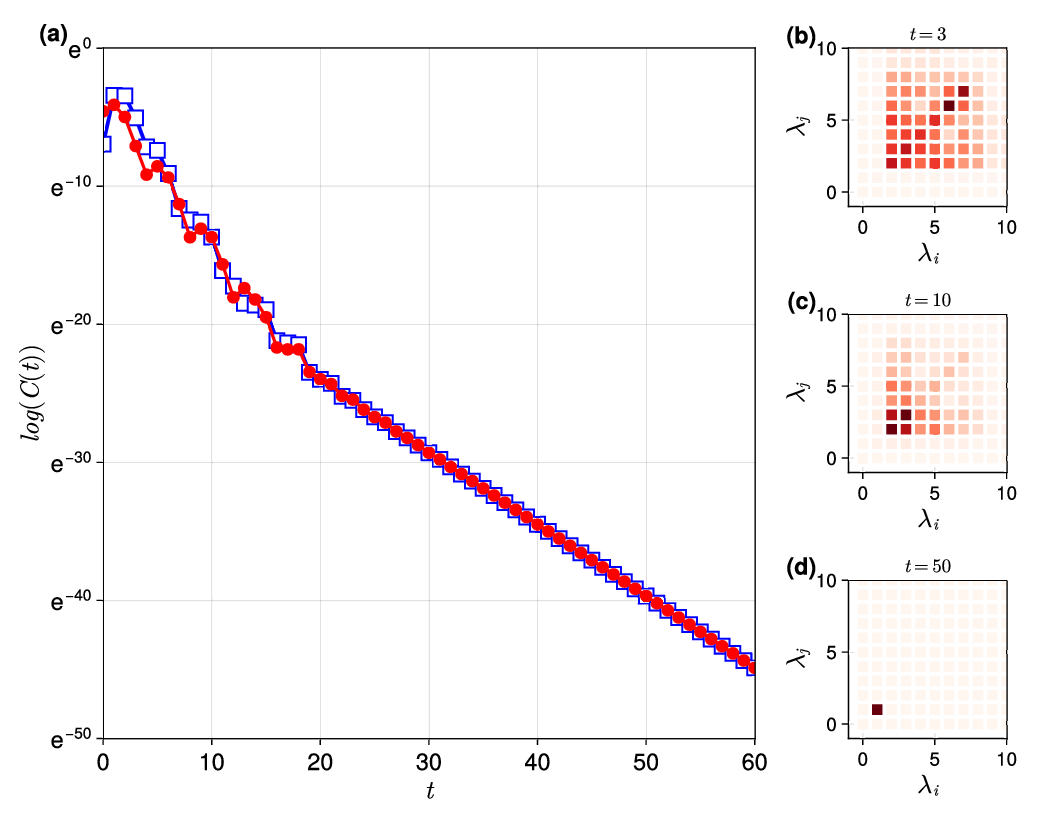}
  \end{center}
  \caption{\textbf{(Color online)} OTOC dynamics and $p_{ij}$ evolution. Panel (A) compares OTOC decay from numerical simulations of \cref{eq:otoc} (blue empty square) with spectral expansion considering only terms from $\lambda_{1}$ to $\lambda_{3}$ (red solid circles) for $K=3.7$. Panels (B)-(D) are heatmaps of $p_{ij}$ magnitude at times $t=3, 10, 50$ respectively, with deeper red/gray indicating larger values. Calculations for Hilbert space dimension $N=1024$, effective Planck constant $h_{eff} = 0.031$ and dissipation rate $\gamma = 0.2$.}
  \label{fig:fig2}
\end{figure}

\begin{figure}[htp]
  \begin{center}
    \includegraphics[width=\columnwidth]{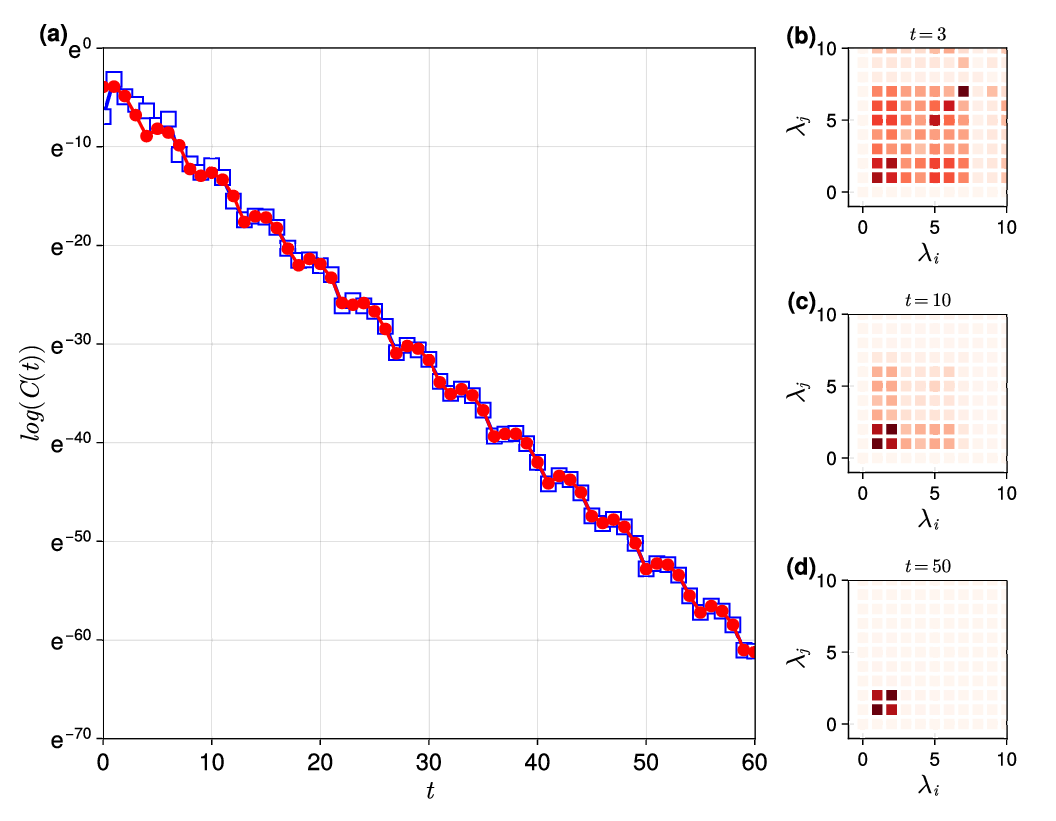}
  \end{center}
  \caption{\textbf{(Color online)} OTOC dynamics and $p_{ij}$ evolution. Panel (A) compares OTOC decay from numerical simulations of \cref{eq:otoc} (blue empty square) with spectral expansion considering only $\lambda_{1}$ and $\lambda_{2}$ (red solid circles) for $K=4.2$. Panels (B)-(D) are heatmaps of $p_{ij}$ magnitude at times $t=3, 10, 50$ respectively, with deeper red/gray indicating larger values. Calculations for Hilbert space dimension $N=1024$, effective Planck constant $h_{eff} = 0.031$ and dissipation rate $\gamma = 0.2$.}
  \label{fig:fig3}
\end{figure}

\begin{figure}[htp]
  \begin{center}
    \includegraphics[width=\columnwidth]{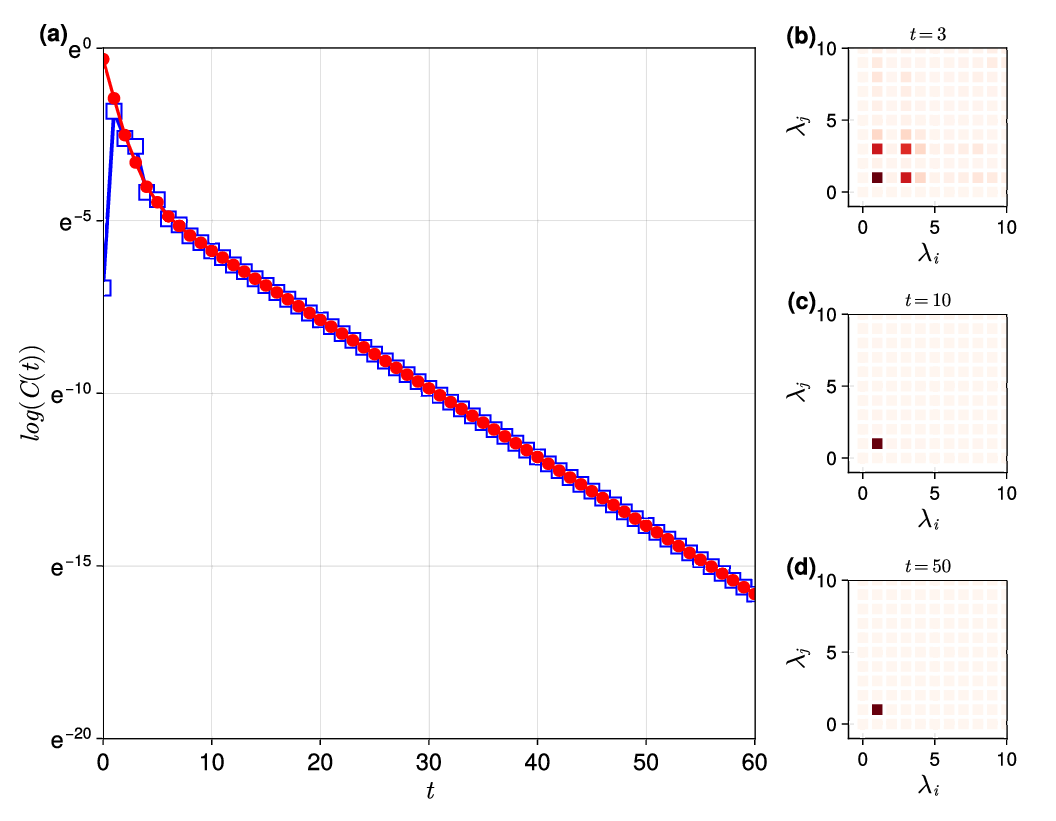}
  \end{center}
  \caption{\textbf{(Color online)} OTOC dynamics and $p_{ij}$ evolution. Panel (A) compares OTOC decay from numerical simulations of \cref{eq:otoc} (blue empty square) with spectral expansion considering only terms from $\lambda_{1}$ to $\lambda_{3}$ (red solid circles) for $K=8.2$. Panels (B)-(D) are heatmaps of $p_{ij}$ magnitude at times $t=3, 10, 50$ respectively, with deeper red/gray indicating larger values. Calculations for Hilbert space dimension $N=1024$, effective Planck constant $h_{eff} = 0.031$ and dissipation rate $\gamma = 0.2$.}
  \label{fig:fig4}
\end{figure}

Figures \cref{fig:fig1,fig:fig2,fig:fig3,fig:fig4} illustrate the results for $K = 2.0$, $3.7$, $4.2$, and $8.2$, corresponding to regimes of regular and chaotic dynamics (see~\cite{bergamasco2023}). Panels~\textit{B}, \textit{C}, and \textit{D} in each figure display the time evolution of $p_{ij}$ for the $10$ largest-modulus eigenvalues at $t = 3$, $10$, and $50$, respectively. Our analysis demonstrates that the spectral development of the OTOC is increasingly dominated by eigenvalues of larger modulus as time progresses. At late times ($t \to 50$), the contribution converges to the leading eigenvalue $\lambda_1$ (and its complex conjugate, if applicable), confirming its dominance in the long-time limit.

At intermediate timescales, however, the contribution of the leading eigenvalue $\lambda_1$ alone proves insufficient to fully capture the OTOC's behavior. This necessitates the inclusion of additional eigenvalues in the spectral decomposition.  

As shown in Panel~\textit{A} of \cref{fig:fig1,fig:fig2,fig:fig3,fig:fig4}, we compare the exact OTOC, computed numerically via Eq.~\ref{eq:otoc}, with its spectral reconstruction using only $\lambda_{1}$ and $\lambda_{2}$. For $K=3.7$ (\cref{fig:fig2}) and $K=8.2$ (\cref{fig:fig4}), we also include the $\lambda_{3}$ terms. Remarkably, even with fewer than four eigenvalues, the reconstructed OTOC retains sufficient fidelity to resolve its decay dynamics.

In the chaotic regimes corresponding to $K = 2.0$, $K = 3.7$, and $K = 4.2$ (see \cref{fig:fig1,fig:fig2,fig:fig3}), where the classical counterpart exhibits chaos, the OTOC reaches values several orders of magnitude smaller than in the regular regime at $K = 8.2$ (see \cref{fig:fig4}). These findings, along with those reported in \cite{bergamasco2023}, highlight that a limited number of propagator eigenvalues play a crucial role in determining the OTOC decay rate and serve as a reliable indicator of the dynamical regime of the system.

\section{\label{sec:conclusion} Conclusion}
In this work, we have explored the spectral decomposition of out-of-time-ordered correlators (OTOCs) in dissipative quantum systems, focusing on their long-time and intermediate-time behavior. By analyzing the eigenvalue spectrum of the quantum Liouvillian, we demonstrated that the OTOC decay can be effectively described using a limited number of dominant eigenvalues, even in cases where the full Liouvillian spectrum is computationally intractable.

Our results reveal two distinct temporal regimes: (i) a long-time decay phase dominated by the spectral gap of the Liouvillian and (ii) an intermediate-time regime where a small subset of subdominant eigenvalues plays a crucial role. This spectral truncation approach provides an efficient method to model OTOC dynamics and highlights a direct connection between quantum dissipation and classical chaos indicators.

Future work could extend this spectral approach to more complex many-body systems and investigate the role of different dissipation mechanisms. Additionally, exploring the experimental feasibility of measuring OTOCs in open quantum platforms would be an important next step in validating these theoretical insights.

\begin{acknowledgments}
Financial support from CONICET and CNEA is gratefully acknowledged.
\end{acknowledgments}

\appendix
\appendix
\section{\label{ap:spectralOTOC} Spectral Decomposition of the OTOC}

Before starting the spectral decomposition of the OTOC, we first establish a simple property of the Lindblad equation. From \cref{eq:lindblad_adj}, it follows that  
\begin{gather}
    \left( \frac{d\hat{B}}{dt} \right)^{\dagger} = \mathcal{L}^{\dagger}(\hat{B}^{\dagger}) = \frac{d\hat{B}^{\dagger}}{dt}
\end{gather}
This result shows that taking the adjoint of an evolving operator is equivalent to evolving the operator first and then applying the transpose and conjugate operations.  

Combining this property with \cref{eq:evolucion_oper} and \cref{eq:descomp_Lambda}, the spectral decomposition of any operator $\hat{B}$ at time $t$ can be written as  
\begin{gather}    
    \hat{B}(t) = (\hat{\Lambda}^{\dagger})^{t} \hat{B}(0) =  \sum_{i}{ (\lambda_{i})^{t}  \langle\braket{L_{i}}{B(0)}\rangle} \ket{R_{i}}\rangle \\
    \hat{B}^{\dagger}(t) = \left((\hat{\Lambda}^{\dagger})^{t} \hat{B}(0)\right)^{\dagger}  =  \sum_{i}{ (\lambda^{*}_{i})^{t}  \langle\braket{L_{i}}{B(0)}\rangle}^{*} \langle\bra{R_{i}}
\end{gather}
Applying the inner product definition, $\langle \braket{A}{B} \rangle = \text{Tr}(A^{\dagger}B)$, we obtain  
\begin{gather}
    \hat{B}(t) = \sum_{i}{ (\lambda_{i})^{t} \text{ Tr}(L_{i}^{\dagger}\hat{B}(0)) R_{i}} \\
    \hat{B}^{\dagger}(t) = \sum_{i}{ (\lambda^{*}_{i})^{t} \left(\text{ Tr}(L_{i}^{\dagger}\hat{B}(0)) \right)^{*} R_{i}^{\dagger}}
\end{gather}
Here, we return to the matrix representation of the eigenvectors of the evolution operator, using the mappings $\ket{R_{i}}\rangle \rightarrow R_{i}$ and $\langle\bra{R_{i}} \rightarrow R_{i}^{\dagger}$.  

Substituting these expressions into \cref{eq:otoc} and employing commutator properties, the spectral decomposition of the OTOC takes the form  
\begin{gather}
    C(t) = \sum_{i,j}{ (\lambda_{i} \lambda^{*}_{j})^{t}\ b_{i}\ b_{j}^{*}\ d_{i,j} }
\end{gather}
where  
\begin{gather}
    b_{i} = \text{ Tr}(L^{\dagger}_{i}\hat{B}(0)), \\
    d_{i,j} = \text{ Tr}\left( \left[ \hat{A},R_{i} \right] \left[ \hat{A}, R_{j} \right]^{\dagger} \rho_{o}\right)
\end{gather}
The expectation value notation $\expval{\cdot} = \text{ Tr}(\cdot \rho_{o})$ is used, with $\rho_{o}$ denoting the initial state.

\end{document}